\documentclass[12pt]{article}  
\makeatletter
\@addtoreset{equation}{section}
\makeatother

\renewcommand{\title}[1]{\null\vspace{25mm}

\noindent{\Large{\bf #1}}\vspace{10mm}

\noindent {\large By }}
\newcommand{\authors}[1]{\noindent{\large #1}\vspace{3mm}

}
\newcommand{\address}[1]{\noindent #1\vspace{5mm}

}
\renewcommand{\abstract}[1]{\vspace{19mm}

\noindent{\small{\em Abstract.} #1}\vspace{2mm}

}

\usepackage{amssymb}
\usepackage{latexsym}
\usepackage{amsthm}
\theoremstyle{plain}
\newtheorem{theorem}{Theorem}[section]
\newtheorem{definition}[theorem]{Definition}
\newtheorem{lemma}[theorem]{Lemma}
\newtheorem{corrolary}[theorem]{Corrolary}

\newcommand{\uline}{\vrule height.06ex depth.02ex width.6em}
\newcommand{\mvee}{\vee\kern-.69em\uline}

\font\1=cmss10

\begin{document}

\markright{\it Helvetica Physica Acta\/\rm, 
[to be published] (1999)}  

\title{Non-Orthomodular Models for Both
\\[2mm] 
Standard Quantum Logic and Standard Classical Logic: 
\\[2mm] 
Repercussions for Quantum Computers}
\authors{Mladen Pavi\v ci\'c$\footnote{
E-mail: mpavicic@faust.irb.hr; Web page: http://m3k.grad.hr/pavicic}$}  
\address{University of Zagreb,
GF, Ka\v ci\'ceva 26, HR-10000 Zagreb, Croatia.}
\authors{and Norman D.~Megill$\footnote{E-mail: nm@alum.mit.edu}$}
\address{Boston Information Group, 30 Church St., 
Belmont MA 02478, U.~S.~A.}
\abstract{It is shown that propositional calculuses of both quantum and 
classical logics are non-categorical. We find that quantum logic 
is in addition to an orthomodular lattice also modeled by a weakly 
orthomodular lattice and that classical logic is in addition to a 
Boolean algebra also modeled by a weakly distributive lattice. Both
new models turn out to be non-orthomodular. We prove the soundness
and completeness of the calculuses for the models. We also prove 
that all the operations in an orthomodular lattice are five-fold 
defined. In the end we discuss possible repercussions
of our results to quantum computations and quantum computers.}



\vbox to 1cm{\vfill}

\section{Introduction}
\label{sec:intro}

For more than a century it has been taken for granted that
the propositional calculus of classical logic has a Boolean 
algebra (complemented distributive lattice) as its only lattice 
model for which completeness of the logic can be proved and for 
more than half a century it has been taken for granted that an 
orthomodular lattice is the only such model of the propositional 
calculus of quantum logic---the logic of quantum mechanics 
\cite{birk-v-neum}. In this paper we prove that both assumptions 
are incorrect by finding a new lattice model for classical logic 
and another for quantum logic neither of which is orthomodular 
(any distributive lattice is orthomodular). We also show that the 
reason why distributive and orthomodular lattices also model 
classical and quantum logics, respectively, lies in the way their 
completeness proofs have been carried out in the past. We show that 
the proofs contained a hidden statement which introduced the 
property of orthomodularity into not necessarily orthomodular 
Lindenbaum algebras of the logics. This is because mappings of the 
logic to an ortholattice does not turn the lattice into an 
orthomodular one as usually assumed. In particular, the 
orthomodularity law and the distributivity law do not map into the 
corresponding lattice expressions at all: the orthomodularity in 
quantum logic and the distributivity in classical logic when mapped 
into a lattice are valid in a non-orthomodular ortholattice and do 
not have anything to do with making the lattices orthomodular 
\cite{mphpa98} and distributive.

In terms of computability our results mean that, structurally,
a computation and inference of formulas neither in classical nor 
in quantum logic correspond to a computation and inference of 
formulas in their models. This discrepancy has not been noticed so 
far because classical calculations in classical computers and 
classical physics in phase space are based not on classical logic 
proper but on its model, i.e., on its distributive model, a Boolean 
algebra. Also an algebra of two valued (yes and no, 1 and 0) 
propositions of classical logic must be a Boolean one and any 
Boolean algebra can be shown equivalent to a Kolmogorovian 
probability theory (which is therefore another possible model for 
classical logic) \cite{leblanc-book}. As opposed to 
this, quantum algebra which would give a Hilbertian probability
theory as a proper universal language for quantum computers
is still not known. Therefore the first idea is to
rely on quantum logic of elementary input propositions themselves.
However, ascribing yes-no values to all quantum propositions is
precluded by the Kochen-Specker theorem.~\cite{peres} Hence, if
one wanted to build a quantum simulator (a general purpose
quantum computer which would not be limited to particular
algorithms such as Shor's or Grover's \cite{plenio}) one should
first develop a proper quantum computer language, i.e., an
algebra which would enable typing in any many-system Schr\"odinger 
equation and then solve it in a polynomial time by simulating the 
systems the equation describes. The need for such an algebra also
stems from the fact that no operation in quantum logic is
unique: as we show in Sec.~\ref{sec:identities} all the operations,
including the identity, are fivefold defined. And with five
identity operations and no definite values ascribable to measurement
propositions we obviously must seek a new algebraic way of
valuating propositions in order to find, for example, which
of them give the same measurement results.

In terms of the model theory, our result means that neither
classical nor quantum logic are categorical. A formal
system is called categorical (monomorphic) if all its models
are isomorphic with each other. In 1934 Tarski was---in
spite of the G\"odel's results---of the opinion that ``a
non-categorical set of sentences (especially if it is used as
an axiomatic system of a deductive theory) does not give the
impression of a closed and organic unity and does not seem to
determine precisely the meaning of the concepts contained in
it.''\cite{tarski} For, the usual set theories are
non-categorical simply because they are incomplete as a
consequence of G\"odel's theorem. The first-order predicate
calculus with Peano's natural number sequence axioms is
non-categorical and complete. In general, it has been
``proved that no consistent first-order theory which possesses
an infinite model is categorical'' simply because ``each
such theory possesses models of arbitrary power.''
(\cite{f-set}, p.~298) Still, simple propositional calculuses
not endowed with quantifiers and numbers which were complete
were apparently expected to be categorical. Now we prove that
surprisingly even such calculuses can be non-categorical.

The paper is organized as follows. In Sec.~\ref{sec:identities}
we show that there are four quantum identities
($a\equiv_i b,\ i=1,...,4$) in an orthomodular lattice which are
not symmetric and one which is ($a\equiv_5 b$). They all boil down
to the classical identity ($a\equiv_0 b$) in a Boolean algebra.
Nevertheless the following implication
$a\equiv_i b=1\ \Rightarrow\ a=b,\ i=1,...,5$
makes an ortholattice orthomodular. Also $a\equiv_0 b=1\
\Rightarrow\ a=b$  makes an ortholattice distributive.
These results we use in Sec.~\ref{sec:ql-models} where we show
that a logic which does have an orthomodular lattice for its
model is not necessarily orthomodular---it also has a weakly
orthomodular model---and in Sec.~\ref{sec:cl-models} that classical
logic which does have a distributive lattice for its model is
not necessarily distributive: it also has a weakly distributive
model. We give soundness and completeness proofs for all the models.

\section{Asymmetrical Quantum Identities}
\label{sec:identities}

An ortholattice (OL) is an algebra ${\cal L}_{\rm O}=
<{\cal L}_{\rm O}^{\circ},',\cap,\cup>$ such that the
following conditions are satisfied for any
$a,b,c\in {\cal L_{\rm O}}^{\circ}$:

\medskip
L1. \qquad $a\cup b\>=\>b\cup a$

\smallskip
L2. \qquad $(a\cup b)\cup c\>=\>a\cup (b\cup a)$

\smallskip
L3. \qquad $a''\>=\>a$

\smallskip
L4. \qquad $a\cup (b\cup b')\>=\>b\cup b'$

\smallskip
L5. \qquad $a\cup (a\cap b)\>=\>a$

\smallskip
L6. \qquad $a\cap b\>=\>(a'\cup b')'$

\bigskip

An orthomodular lattice OML is an OL
in which the following additional condition is satisfied:

\smallskip
L7. \qquad $a\cup b=((a\cup b)\cap b')\cup b$

\smallskip
A weakly  orthomodular lattice WOML is an OL
in which the following additional condition is satisfied:

\smallskip
L8. \qquad $(a'\cap (a\cup b))\cup b'\cup (a\cap b)=1$

\smallskip
A distributive lattice (Boolean algebra) DL is an
OL in which the following additional condition is satisfied:

\smallskip
L9. \qquad $a\cup(b\cap c)=(a\cup b) \cap (a\cup c)$.

\smallskip It is well-known that in every orthomodular lattice
five polynomial implications satisfy the Birkhoff-von Neumann
requirement \cite{kalmb83}:
\begin{eqnarray}
a\rightarrow_i b=1\qquad \Rightarrow\qquad a\le b, \qquad\qquad
i=1,\dots,5,
\label{eq:von-N-B}
\end{eqnarray}
where $a\rightarrow_1b\ {\buildrel\rm def\over =}\ a'\cup(a\cap b)$,
$a\rightarrow_2b\ {\buildrel\rm
def\over =}\ b'\rightarrow_1a'$, $a\rightarrow_3b\
{\buildrel\rm def\over =}\ (a'\cap b)\cup(a'\cap b')\cup
(a\rightarrow_1b)$, $a\rightarrow_4b\ {\buildrel\rm def\over =}\
b'\rightarrow_3a'$, and $a\rightarrow_5b\ {\buildrel\rm def\over =}\
(a\cap b)\cup(a'\cap b)\cup(a'\cap b')$.

Even more, it can be proved \cite{pav87} that the rule (\ref{eq:von-N-B})
makes an ortholattice orthomodular, i.e., that (\ref{eq:von-N-B}) can
be substituted for L7. Since it can also be proved \cite{p98}
that the following rule
\begin{eqnarray}
a\rightarrow_0 b=1\qquad \Rightarrow\qquad a\le b,
\label{eq:boole}
\end{eqnarray}
where $a\rightarrow_0b\ {\buildrel\rm def\over =}\ a'\cup b$, makes
an ortholattice distributive (Boolean algebra),  i.e., that
(\ref{eq:boole}) can be substituted for L9, it is clear that
$a\rightarrow_i b,\ i=1,\dots,5$ all merge to $a\rightarrow_0 b$ in a
classical theory. In addition, one can write any $a'$ as
$a\rightarrow_i 0$ and one can prove\cite{mpijtp98}:
\begin{eqnarray}
a\cup b=(a\rightarrow_i b)\rightarrow_i(((a
\rightarrow_i b)\rightarrow_i(b \to_i a))\rightarrow_i a)
\label{eq:vel}
\end{eqnarray}
for $i=1,\dots,5$. Thus one can form a quantum implication algebra
with the operation of implication as a single primitive and to
prove that an orthomodular (distributive) lattice can model
quantum (classical) logic seems to be obvious since it is
easy to prove that in any orthomodular
lattice we have: $a\leftrightarrow_i b=a\equiv_5 b,\ i=1,\dots,5$,
where $a\leftrightarrow_i b\ {\buildrel\rm def\over =}\ (a\rightarrow_i
b)\cap(b\rightarrow_i a)$ and $a\equiv_5 b\ {\buildrel\rm def\over =}\
(a\cap b)\cup(a'\cap b')$ and the identity operation $a\equiv_5 b$ reduces
to $a\equiv_0 b\ {\buildrel\rm def\over =}\ (a'\cup b)\cap(b'\cup a)$
in a classical theory. For, $a\equiv_i b=1,\ i=0,5$ is reflexive,
symmetric, and transitive and therefore is a relation of equivalence
and seems applicable for completeness proofs of our logics.

However, the first doubts are raised by the results  that

\begin{eqnarray}
a\equiv_5 b=1\qquad \Rightarrow\qquad a=b,
\label{eq:qm-id}
\end{eqnarray}
makes an ortholattice orthomodular \cite{pav93} and that
\begin{eqnarray}
a\equiv_0 b=1\qquad \Rightarrow\qquad a=b,
\label{eq:cl-id}
\end{eqnarray}
makes an ortholattice distributive \cite{p98}.

A real confirmation of our doubts comes from considering mixed
biimplications. All implications reduce to the classical one in a
classical theory, so, not only $a\leftrightarrow_i b$ but also
$(a\rightarrow_i b)\cap(b\rightarrow_j a)$, $i\ne j$ must reduce
to $a\equiv_0 b$ in a classical theory. Let us have a look at what
we get in an orthomodular lattice in Table 1,
where $a\equiv_1 b\ {\buildrel\rm def\over =}\ (a\cup b')\cap
(a'\cup (a\cap b))$, $a\equiv_2 b\ {\buildrel\rm def\over =}\ (a\cup b')\cap
(b\cup (a'\cap b'))$, $a\equiv_3 b\ {\buildrel\rm def\over =}\ (a'\cup b)\cap
(a\cup (a'\cap b'))$ and $a\equiv_4 b\ {\buildrel\rm def\over =}\
(a'\cup b)\cap(b'\cup (a\cap b))$.
We omit the easy proof. We can also send the reader a computer
program which reduces any two-variable orthomodular lattice expression
to one of the 96 simplest possible ones as given in \cite{beran}.

\begin{tabular}{|lr||c|c|c|c|c|c|} \multicolumn{8}{c}{}\\ \hline
$\ ^i\  _\downarrow \ ${\huge$\backslash$} & $^j\ _\rightarrow$
& $b\rightarrow_0 a$ & $b\rightarrow_1 a$ & $b\rightarrow_2 a$ &
$b\rightarrow_3 a$ & $b\rightarrow_4 a$ & $b\rightarrow_5 a$
\\ \hline \hline
\multicolumn{2}{|c||}{$a\rightarrow_0b$} &\ \ $a\equiv_0 b\ \ $
 &\ \  $a\equiv_4 b\ \ $  &\ \  $a\equiv_3 b\ \ $  &
\  \ $a\equiv_2 b\ \ $  &\  \ $a\equiv_1 b\ \ $
 &\ \  $a\equiv_5 b\ \ $ \\ \hline
\multicolumn{2}{|c||}{$a\rightarrow_1b$} & $a\equiv_1 b$ & $a\equiv_5 b$
& $a\equiv_5 b$ &
$a\equiv_5 b$ & $a\equiv_1 b$ & $a\equiv_5 b$\\ \hline
\multicolumn{2}{|c||}{$a\rightarrow_2b$} & $a\equiv_2 b$ &
$a\equiv_5 b$ & $a\equiv_5 b$ &
$a\equiv_2 b$ & $a\equiv_5 b$ & $a\equiv_5 b$\\ \hline
\multicolumn{2}{|c||}{$a\rightarrow_3b$} & $a\equiv_3 b$ &
$a\equiv_5 b$ & $a\equiv_3 b$ &
$a\equiv_5 b$ & $a\equiv_5 b$ & $a\equiv_5 b$\\ \hline
\multicolumn{2}{|c||}{$a\rightarrow_4b$} & $a\equiv_4 b$ &
$a\equiv_4 b$ & $a\equiv_5 b$ &
$a\equiv_5 b$ & $a\equiv_5 b$ & $a\equiv_5 b$\\ \hline
\multicolumn{2}{|c||}{$a\rightarrow_5b$} & $a\equiv_5 b$ &
$a\equiv_5 b$ & $a\equiv_5 b$ &
$a\equiv_5 b$ & $a\equiv_5 b$ & $a\equiv_5 b$\\ \hline
\multicolumn{8}{c}{}\\
\multicolumn{8}{c}{Table 1: Products $(a\rightarrow_i b)
\cap(b\rightarrow_j a)$, $i=0,\dots,5$ (rows), $j=0,\dots,5$
(columns).}\\
\multicolumn{8}{c}{ ``Identities'' $a\equiv_i b,\ i=1,\dots,4$ are
asymmetrical.} \\
\multicolumn{8}{c}{}\\
\end{tabular}

\begin{figure}[htbp]\centering
  \setlength{\unitlength}{1pt}
  \begin{picture}(260,150)(-10,-10)

    \put(70,0) { 
      \begin{picture}(124,120)(0,0) 
        \put(40,0){\line(-1,1){40}}
        \put(40,0){\line(0,1){40}}
        \put(40,0){\line(1,1){40}}
        \put(00,40){\line(0,1){20}}
        \put(00,40){\line(1,1){40}}
        \put(40,40){\line(-1,1){40}}

        \put(40,40){\line(1,1){40}}
        \put(80,40){\line(-1,1){40}}
        \put(80,40){\line(0,1){20}}
        \put(00,60){\line(0,1){20}}
        \put(80,60){\line(0,1){20}}
        \put(00,80){\line(1,1){40}}
        \put(40,80){\line(0,1){40}}
        \put(80,80){\line(-1,1){40}}
        \put(10,70){\line(-1,-3){10}}
        \put(70,50){\line(1,3){10}}

        \put(40,-5){\makebox(0,0)[t]{$0$}}
        \put(-5,40){\makebox(0,0)[r]{$x$}}
        \put(40,45){\makebox(0,0)[b]{$w$}}
        \put(85,40){\makebox(0,0)[l]{$z'$}}
        \put(-5,60){\makebox(0,0)[r]{$y$}}
        \put(85,60){\makebox(0,0)[l]{$y'$}}
        \put(-5,80){\makebox(0,0)[r]{$z$}}
        \put(40,73){\makebox(0,0)[t]{$w'$}}
        \put(85,80){\makebox(0,0)[l]{$x'$}}
        \put(40,125){\makebox(0,0)[b]{$1$}}
        \put(15,70){\makebox(0,0)[l]{$v$}}
        \put(65,50){\makebox(0,0)[r]{$v'$}}

        \put(40,0){\circle*{3}}
        \put(00,40){\circle*{3}}
        \put(40,40){\circle*{3}}
        \put(80,40){\circle*{3}}
        \put(00,60){\circle*{3}}
        \put(80,60){\circle*{3}}
        \put(00,80){\circle*{3}}
        \put(40,80){\circle*{3}}
        \put(80,80){\circle*{3}}
        \put(40,120){\circle*{3}}
        \put(10,70){\circle*{3}}
        \put(70,50){\circle*{3}}
      \end{picture}
    } 

  \end{picture}
  \caption{
     Ortholattice
M12.
\label{fig-mccune2}}
\end{figure}
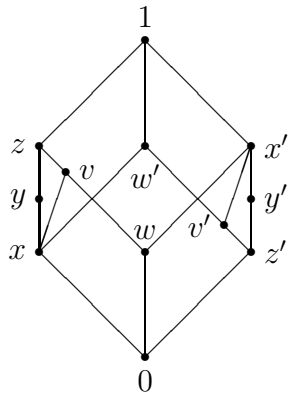

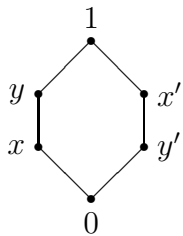
\begin{figure}[htbp]\centering
  \begin{picture}(60,80)(-10,-10)

    \put(20,0){\line(-1,1){20}}
    \put(20,0){\line(1,1){20}}
    \put(0,20){\line(0,1){20}}
    \put(40,20){\line(0,1){20}}
    \put(0,40){\line(1,1){20}}
    \put(40,40){\line(-1,1){20}}

    \put(20,-5){\makebox(0,0)[t]{$0$}}
    \put(-5,20){\makebox(0,0)[r]{$x$}}
    \put(45,20){\makebox(0,0)[l]{$y'$}}
    \put(-5,40){\makebox(0,0)[r]{$y$}}
    \put(45,40){\makebox(0,0)[l]{$x'$}}
    \put(20,65){\makebox(0,0)[b]{$1$}}

    \put(20,0){\circle*{3}}
    \put(0,20){\circle*{3}}
    \put(40,20){\circle*{3}}
    \put(0,40){\circle*{3}}
    \put(40,40){\circle*{3}}
    \put(20,60){\circle*{3}}

  \end{picture}
\caption{Ortholattice O6\label{fig:O6}}
\end{figure}

Also, we are able to prove:
\begin{lemma}In any {\em OML} we have:
\begin{eqnarray}
a\equiv_i b=(a\rightarrow_i b)\cap(b\rightarrow_0
a)\qquad\qquad\qquad\qquad
i=0,\dots,5.\label{eq:id-impl}
\end{eqnarray}
This also holds in any {\em OL} for $i=0,1,2$ and in some {\em OL}s
weaker than {\em OML} for $i=3,4,5$.
\end{lemma}
\begin{proof}
We omit the easy proof that Eq.~(\ref{eq:id-impl}) holds in any
OML.  For $i=0,1,2$ that it holds in any OL is apparent from the
definitions.  For $i=3,4,5$ it fails in the non-orthomodular
ortholattice from Fig.~\ref{fig-mccune2}\footnote{The
authors would like to thank to
William McCune, Argonne National Lab, Argonne IL, U.~S.~A.
({\tt http://www.mcs.anl.gov/home/mccune/ar/ortholattice/}), for
finding this lattice, using the matrix-finding program MACE.}
but does not fail either in O6 (Fig.~\ref{fig:O6}) or in
WOML, non-OML lattices from \cite{beran}, Figs.~7b, 9f, 9h, and 11.
\end{proof}

The expressions $a\equiv_i b$, $=1,\dots,4$ are all asymmetrical
and at first we would think it would be inappropriate to name
them identities. But we are able to prove the following theorem.

\medskip\noindent
\begin{theorem}\label{th:other-eq}An ortholattice in which
\begin{eqnarray}
a\equiv_i b=1\qquad \Rightarrow\qquad a=b, \qquad\qquad
i=1,\dots,4\label{eq:qm-as-id}
\end{eqnarray}
holds is an orthomodular lattice and vice versa.
\end{theorem}

\begin{proof}We give here the proof only for $i=1$.
Others are completely analogous. Let us write the
premise $a\equiv_1 b=1$ as $(a\cup b')\cap(a\rightarrow_1b)=1$.
Hence, $(a\rightarrow_1b)=1$ and according to \cite{pav89}
$a\le b$. This, together with the other consequence of the
premise: $(a\cup b')=1$, yields $b\le a$ \cite{pav89}, what
proves the statement.

As for the vice versa part, all four implications fail in O6 which 
means that they must be orthomodular.
\end{proof}

%

\medskip
Hence, putting together Eq.~(\ref{eq:qm-id}) and
Eq.~(\ref{eq:qm-as-id}) we have an indication that the
relation of equivalence which establishes a connection between
quantum logic and its models might turn out to be based on several
different operations of identity at the same time thus making
a direct evaluation of elementary logical propositions
impossible. In Sec.~\ref{sec:ql-models} we  prove the conjecture.
In Sec.~\ref{sec:cl-models} we analyze classical logic and show
that although its syntactical structure can map into a weakly
distributive lattice as a model, properties of the Boolean
algebra as another model enable a consistent direct evaluation
of elementary logical propositions.

\section{Non-Orthomodular Model for Quantum Logic}
\label{sec:ql-models}

A reader which is not at home with methods and parlance
of mathematical logic can follow this section by reading
logical expressions of the form $\vdash A$ as $a=1$ in the
lattice language. In doing so he or she will miss some features of
a proper logic as, for example, that in a logic $a\wedge b$ and
$b\wedge a$ are distinct formulas (they coincide in a lattice)
but these features do not play an important role in our proofs.
We are only interested in connecting the equivalence relations
in our logic---which coincide with those in a lattice---with equations
in lattices.

A crucial difference we find between quantum logic and
orthomodular lattice as its standard model is that properties that
play a decisive role in the lattice do not play such a role in the
logics. This is in contrast with the properties of our new model,
weakly orthomodular lattice, whose properties do correspond to
those of the logic. To explain these differences let us consider the
orthomodularity property. When we add the orthomodularity property
to an ortholattice it becomes an orthomodular lattice. We can compare
what happens in a logic by looking at a lattice we obtain by mapping
logical axioms $\vdash A$ to an ortholattice where they take over
the form $a=1$; here $a=f(A)$ and $f$ is a morphism from the logic
to the lattice. As we have shown in \cite{mphpa98} the property
$(a\cup(a'\cap(a\cup b)))\equiv_5(a\cup b)=1$,
we obtain by mapping the logical formula for ``orthomodularity''
$\vdash (A\vee(\neg A\wedge(A\vee B))\equiv_5(A\vee B)$ into an
ortholattice, is true in all ortholattices. The reason for such
different structures of logic as opposed to its standard  model lies
in the way we prove the completeness of the standard modeling.
To understand this better we give both completeness proofs:
in Subsection \ref{subsec:q-compl-st} for the standard model
and in \ref{subsec:woml-complete} for the new one.

We first consider a quantum logic ($\cal QL$)  derived directly
from the properties of a weakly orthomodular lattice WOML without
taking the orthomodularity property into account.
We do so in order to show that orthomodularity appears only at the
stage of proving the completeness and as a property of equivalence
classes we can define on a logic.  $\cal QL$ is equivalent to the
logics of other authors, e.~g., Kalmbach's \cite{kalmb83},
Dishkant's \cite{dishk} , Dalla Chiara's \cite{dalla-c-h-b} ,
Mittelstaedt's \cite{mittelstaedt-book},
Stachow's \cite{stachow-completness}, Hardegree's \cite{harde79},
R\"uttimann's \cite{ruettimann-book}, etc.
We proved explicitly the equivalence to Kalmbach's and Dishkant's
systems in \cite{mphpa98} but a general equivalence to all systems
follows from our completeness proof given below.

\subsection{Quantum Logic}
\label{subsec:q-l}

Quantum logic $\cal QL$ contains the connectives $\rightarrow$,
$\leftrightarrow$, $\equiv$, $\vee$, $\wedge$, and $\neg$
which we represent with their lattice counterparts:
$\rightarrow$, $\leftrightarrow$, $\equiv$, $\cup$,
$\cap$, and $'$. Let ${\cal F}^\circ$ be the set
of all logical expressions, i.e., well formed formulas (wff).
Of these $\vee$, $\neg$ and $\cup$, $'$ are primitive ones. 
The latter constitutes an algebra 
${\cal F}=\langle {\cal F}^\circ,\neg,\vee,\wedge\rangle$.
$\cal QL$ is given by the following axioms and rules
of inference, representing five distinct but equivalent
systems.

\bigskip
\noindent
{\bf Axioms}

\medskip
{\bf $\cal QL$1.}\quad\qquad$\vdash A\vee B\equiv_i B\vee A$

\smallskip
{\bf $\cal QL$2.}\quad\qquad$\vdash A\vee (B\vee C)\equiv_i (A\vee B)\vee C$

\smallskip
{\bf $\cal QL$3.}\quad\qquad$\vdash A\equiv_i\neg\neg A$

\smallskip
{\bf $\cal QL$4.}\quad\qquad$\vdash\neg A\vee A\equiv_i(\neg A\vee A)\vee B $

\smallskip
{\bf $\cal QL$5.}\quad\qquad$\vdash A\vee (A\wedge B)\equiv_i A$

\smallskip
{\bf $\cal QL$6.}\quad\qquad$\vdash(A\wedge B)\equiv_i\neg(\neg A\vee\neg B)$

\smallskip
\noindent
where $i=1,\dots,5$ and will be considered to take over a specific
value
throughout.

\medskip\noindent
{\bf Rules of Inference}

\medskip
{\bf $\cal QL$R1.}\qquad$\vdash A\equiv_iB
\qquad\Rightarrow\qquad\vdash A\vee C\equiv_i B\vee C$

\medskip
{\bf $\cal QL$R2.}\qquad$\vdash A\equiv_iB
\qquad\&\qquad\vdash B\equiv_i C
\qquad\Rightarrow\qquad\vdash A\equiv_i C$

\medskip
{\bf $\cal QL$R3.}\qquad$\vdash A\equiv_iB
\qquad\Leftrightarrow\qquad\vdash \neg A\equiv_i \neg B$

\medskip
{\bf $\cal QL$R4.}\qquad$\vdash A\equiv_iB
\qquad\Rightarrow\qquad\vdash B\equiv_i A$

\medskip
{\bf $\cal QL$R5.}\qquad$\vdash \neg A\vee A\equiv_iB
\qquad\Leftrightarrow\qquad\vdash B$

Axioms $\cal QL$1-6 coincide with L1-6 of OL, and
$\cal QL$R1 with L8 (in the form of L8.1 shown
in Theorem \ref{th:l8.1} below).

\begin{definition}\label{D:gamma-ql}
For $\Gamma\subseteq {\cal F}^\circ$ we say $A$ is derivable from
$\Gamma$ and write $\Gamma\vdash A$ if there is a sequence of
formulas ended by $A$ each of which is either one of the axioms of
$\cal QL$ or is a member of $\Gamma$ or is obtained from its
precursors with the help of a rule of inference of the logic.
\end{definition}

\begin{definition}\label{model-ql}We call ${\cal M}=\langle
{\cal L},f\rangle$ a model of a set of formulas  $\Gamma$,
if ${\cal L}$ is a lattice {\em (WOML} or {\em OML)},
$f:{\cal F}^\circ\longrightarrow {\cal L}$
is a morphism of algebra of wff's which satisfies $f(A)=1$ for
any $A\in\Gamma$; we call the latter $A$ true in the model
${\cal M}$.
\end{definition}

\subsection{Soundness Proof for Quantum Logic}
\label{subsec:q-sound}

Lemmas \ref{le:ol}--\ref{le:teq1ded} provide some technical results
for use in subsequent proofs.

\begin{lemma}\label{le:ol} In any {\em OL} we have:
\begin{eqnarray}
&(a \cap b) \cup (a \cap c) \le a \cap(b \cup c) \label{eq:lem-halfdist}\\
&a=1 \qquad\&\qquad a\to_0 b=1 \qquad\Rightarrow\qquad b=1\label{eq:lem-mp}\\
&(a\equiv_5 b)\to_0(a\leftrightarrow_1 b)=1\label{eq:lem-bibi}\\
&b\to_2 a=1 \qquad\Rightarrow\qquad a \to_2 (a \equiv_5 b) = a \equiv_5 b
  \label{eq:wlem3a}\\
&a\equiv_5 b=1 \qquad\Rightarrow\qquad a\to_1 (b\cup c)=1\label{eq:wlem4b}\\
&a \to_2(b \cup c)=(a\cup c)\to_2(b\cup c)\label{eq:wlem4a}\\
&a\to_i (a \cap b)\ =\ a\equiv_i(a\cap b)\ =\ (a \cap b)\equiv_i a\ =\ a\to_1 b
  \qquad i=0,\ldots,5\quad
  \label{eq:i1} \\
&(a \cup b)\to_i b\ =\ (a\cup b)\equiv_i b\ =\ b\equiv_i(a\cup b)\ =\ a\to_2 b
  \qquad i=0,\ldots,5\quad
  \label{eq:i2}
\end{eqnarray}
\end{lemma}
\begin{proof}
For (\ref{eq:lem-halfdist}): This is well known and we omit the proof.
For (\ref{eq:lem-mp}): See \cite[p.~237]{kalmb83}.
For (\ref{eq:lem-bibi}):
$a\equiv_5 b \le a \to_1 b$ and
$\le b \to_1 a$, so
$a\equiv_5 b \le (a \to_1 b) \cap (b \to_1 a)$;
$1 = (a \equiv_5 b)' \cup (a \equiv_5 b) \le
(a\equiv_5 b)' \cup ((a\to_1 b)\cap(b\to_1 a))$.
For (\ref{eq:wlem3a}):  From L5 and DeMorgan's law we have
$a' \cap ( a' \cup b')=a'$, so
$a' \cap ( a' \cup b') \cap (a \cup b)=a'\cap(a \cup b)$;
from hypothesis
and DeMorgan's we have
$a' \cap(a \cup b)=0$, so
$a' \cap ( a' \cup b') \cap (a \cup b)=0$; from DeMorgan's we have
$(a \equiv_5 b)' = (a'\cup b')\cap (a \cup b)$, so
$a' \cap (a \equiv_5 b)' = 0$, so
$(a \equiv_5 b) \cup (a' \cap (a \equiv_5 b)') = a \equiv_5 b$.
For (\ref{eq:wlem4b}): From hypothesis and (\ref{eq:lem-bibi})
and (\ref{eq:lem-mp}) we have $a\leftrightarrow_1 b=1$, so
$1=a\leftrightarrow_1 b\le a\to_1 b =a'\cup(a\cap b) \le
  a'\cup(a\cap(b\cup c)) = a\to_1(b\cup c)$.
For (\ref{eq:wlem4a}):
$a'\cap(b\cup c)'=(a'\cap b')\cap c'=(a'\cap c')\cap(b'\cap c')
=(a\cup c)'\cap(b\cup c)'$, so
$(b\cup c)\cup(a'\cap(b\cup c)')=(b\cup c)\cup((a\cup c)'\cap(b\cup c)')$.
For (\ref{eq:i1}) and (\ref{eq:i2}): We omit the easy verifications.
\end{proof}

\begin{lemma}In any {\em WOML} we have:
\begin{eqnarray}
&(a\to_1 b)\to_0(a \to_2b)=1\label{eq:l8a}\\
&a\to_1 b=1 \qquad\Leftrightarrow\qquad a\to_2 b=1\label{eq:wlem1} \\
&a\to_2 b=1 \qquad\Rightarrow\qquad a \to_2 (a \equiv_5 b)=1
  \label{eq:wlem3b}\\
&a\to_2 b=1 \qquad\&\qquad b\to_2 a=1 \qquad\Rightarrow\qquad a\equiv_5b=1
  \label{eq:wlem3}\\
&a\equiv_5 b=1 \qquad\Rightarrow\qquad a\to_2 (b\cup c)=1\label{eq:wlem4c}\\
&a\equiv_5 b=1 \qquad\Rightarrow\qquad (a\cup c)\equiv_5(b\cup c)=1
  \label{eq:wlem4}
\end{eqnarray}
\end{lemma}
\begin{proof}
For (\ref{eq:l8a}): Immediate from L8, L1, L3 and
definitions.
For (\ref{eq:wlem1}):
Immediate from L8 and (\ref{eq:l8a}), using (\ref{eq:lem-mp}).
For (\ref{eq:wlem3b}): Using (\ref{eq:lem-halfdist}),
$a \cap b = a \cap(a \cap b) \le (a\cap(a \cap b))\cup (a\cap(a' \cap b'))
   \le a \cap((a\cap b)\cup(a' \cap b'))= a\cap(a \equiv_5 b)$; so
$a\to_1 b = a' \cup(a\cap b) \le a' \cup(a \cap(a \equiv_5 b))
   = a \to_1 (a \equiv_5 b)$;
so from hypothesis and (\ref{eq:wlem1}) we have
$1 = a\to_1 b \le a \to_1 (a \equiv_5 b)$; so from (\ref{eq:wlem1})
we conclude
$1 = a \to_2 (a \equiv_5 b)$.
For (\ref{eq:wlem3}): Immediate from (\ref{eq:wlem3a}) and
(\ref{eq:wlem3b}).
For (\ref{eq:wlem4c}): Immediate from (\ref{eq:wlem4b}) and
(\ref{eq:wlem1}).
For (\ref{eq:wlem4}): From (\ref{eq:wlem4c}) we have $a\to_2 (b\cup c)=1$
and $b\to_2 (a\cup c)=1$; so from (\ref{eq:wlem4a}) we have
 $(a\cup c)\to_2(b\cup c)=1$ and $(b\cup c)\to_2(a\cup c)=1$; so from
(\ref{eq:wlem3}) we have $(a\cup c)\equiv_5(b\cup c)=1$.
\end{proof}

\begin{lemma}\label{le:teq1}
Let $t$ be any term (such as $a\cup a'$).  If the equation
$t=1$ holds in all {\em OML}s, then $t=1$ holds in {\em OL
+ (\ref{eq:wlem4})}.
\end{lemma}
\begin{proof}
Theorem 2.15 in \cite{mphpa98} and the remark after Theorem 2.12 in
\cite{mphpa98}, which applies to any OL in which (\ref{eq:wlem4}) holds.
\end{proof}

\begin{lemma}\label{le:oldwom}(a) An ortholattice in which {\em 
(\ref{eq:wlem4})} holds is a {\em WOML} and vice versa. (b) An 
ortholattice in which either direction of {\em (\ref{eq:wlem1})} 
holds is a {\em WOML} and vice versa.
\end{lemma}
\begin{proof}
(a) We have shown that (\ref{eq:wlem4}) holds in a WOML.  On the other
hand, it is easy to prove (using the Foulis-Holland theorem for example)
that $(a'\cap (a\cup b))\cup b'\cup (a\cap b)=1$ holds in an OML; thus
by Lemma~\ref{le:teq1} it also holds in OL + (\ref{eq:wlem4}).  In other
words, the WOML we have defined here is equivalent to the WOML of
\cite{mphpa98}, and L8 and (\ref{eq:wlem4}) are interchangeable as the
WOM law added to an OL.  (b) It is easy to prove either direction of
(\ref{eq:wlem1}) from the other using only L1--L6.  In the proof of
(\ref{eq:wlem4}), we used only (\ref{eq:wlem1}) along with L1--L6.
Thus in an OL, (\ref{eq:wlem1}) follows from L8, and
(\ref{eq:wlem4}) follows from (\ref{eq:wlem1}).
\end{proof}

\begin{lemma}\label{le:teq1ded}
Let $t_1,...,t_n,t$ be any terms ($n\ge 0$).  If the inference
$t_1=1\ \&\ \ldots$ $\&\ t_n=1\ \Rightarrow\ t=1$ holds
in all {\em OML}s, then it holds in any {\em WOML}.
\end{lemma}
\begin{proof}
We extend the proof of Theorem 2.12 of \cite{mphpa98} using the
completeness proof for unary quantum logic (e.g.~\cite[p. 238]{kalmb83})
where $t_1=1,\ \ldots,\ t_n=1$ are the ortholattice mappings for
the hypotheses of a deduction.
\end{proof}

\begin{theorem}\label{th:l8.1} {\em WOML} given as {\em L1--L6 + L8
[}which can also be written as $(a\rightarrow_1 b)\rightarrow_0(a
\rightarrow_2b)=1${\em ]} is an {\em OL} to which the following
mapping of $\cal QL${\em R1}

\smallskip
{\em L8.1}. \qquad$a\equiv_ib=1\qquad\Rightarrow\qquad
a\cup c\equiv_i b\cup c=1\qquad\qquad i=1,\ldots,5$

\smallskip\noindent
is added and vice versa.
\end{theorem}
\begin{proof}
Since $a=b$ implies $a\cup c=b\cup c$, by Theorem~\ref{th:other-eq}
and (\ref{eq:qm-id}) we have that
$a\equiv_ib=1$ implies
$(a\cup c)\equiv_i (b\cup c)=1$ in any OML.
By Lemma~\ref{le:teq1ded} this also holds in any WOML.
On the other hand, assume L8.1 holds.  If
$a\to_1 b=1$ then $a\equiv_i (a\cap b)=1$
by (\ref{eq:i1}), so $(a\cup b)\equiv_i ((a\cap b)\cup b)=1$ by L8.1, so
$(a \cup b)\equiv_ib=1$, so $a\to_2 b=1$ by (\ref{eq:i2}), so L8
holds by Lemma~\ref{le:oldwom}b.
\end{proof}

Let us also prove the following theorem which we shall use
later on.
\begin{theorem} {\em WOML} is an {\em OL}
to which either of the following properties is added:

\smallskip
{\em L8.2}. \qquad$a\rightarrow_1b=1\qquad\Rightarrow\qquad
b'\rightarrow_1 a'=1$

\smallskip
{\em L8.3}. \qquad$((a\rightarrow_1b)\rightarrow_0 b)\ \equiv_5\
(a\cup b)=1$

\smallskip\noindent
and vice versa.
\end{theorem}

\begin{proof}
For L8.2:  Immediate from Lemma~\ref{le:oldwom}b.
For L8.3: We have, using L1--L6,
$(a\to_1 b)\to_0 b = (a\cap(a'\cup b'))\cup b \le a\cup b$, so
$((a\to_1 b)\to_0 b) \cap(a\cup b)=(a\to_1 b)\to_0 b$ and
$((a\to_1 b)\to_0 b)' \cap(a\cup b)'=(a\cup b)'$.  Hence
$((a\to_1 b)\to_0 b)\equiv_5 (a\cup b)=((a\to_1 b)\to_0 b)\cup (a\cup b)'
=(a\cap(a'\cup b'))\cup b\cup (a' \cap b')$, which becomes the
left-hand side of L8 after substituting $a'$ for $a$ and
$b'$ for $b$ then applying L3.
\end{proof}

We now prove the soundness of modeling quantum logic by
a weakly orthomodular lattice.

\begin{theorem}\label{th:soundness}(Soundness)
If \ $\Gamma\vdash A$, then $A$ is true in any {\em WOML} model.
\end{theorem}

\begin{proof}
Any axiom $\cal QL$1---$\cal QL$6 is true in any model
WOML. Let us put $a=f(A)$ and $b=f(B)$ and let
us verify for example $\cal QL$1 for $i=5$. It maps to
$((a\cup b)\cap(b\cup a))\cup
((a'\cap b')\cap(b'\cap a'))=1$. By L1, L2, L4, and L8 we get
$(a\cup b)\cup(a\cup b)'=1$ which is true by definition.
$\cal QL$2---$\cal QL$6 we prove analogously.
We also have to verify that the set of formulas true in a
model ${\cal M}$ is closed under the
rules of inference: $\cal QL$R1-5. $\cal QL$R1 maps to
L8.1. $\cal QL$R2 maps to $a\equiv_ib=1\ \&\ b\equiv_i c
=1\ \Rightarrow\ a\equiv_ic=1$ which according to
Lemma~\ref{le:teq1ded}, Theorem~\ref{th:other-eq} and
(\ref{eq:qm-id}) holds in any WOML since $a=b\ \&\ b=c\
\Rightarrow a=c$ holds in any OML. $\cal QL$R3-5 mappings
we verify analogously.
 \end{proof}

\subsection{Standard Completeness Proof for Quantum Logic}
\label{subsec:q-compl-st}

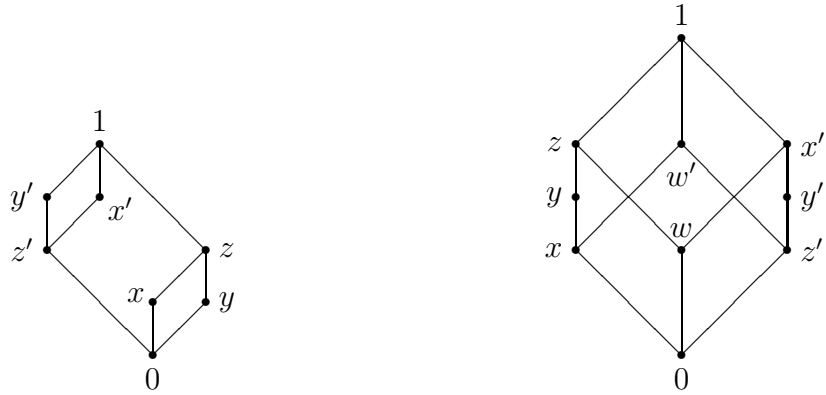
\begin{figure}[htbp]\centering
  \begin{picture}(260,150)(-10,-10)

    \put(40,0){\line(-1,1){40}}
    \put(40,0){\line(0,1){20}}
    \put(40,0){\line(1,1){20}}
    \put(40,20){\line(1,1){20}}
    \put(60,20){\line(0,1){20}}
    \put(0,40){\line(0,1){20}}
    \put(0,40){\line(1,1){20}}
    \put(60,40){\line(-1,1){40}}
    \put(0,60){\line(1,1){20}}
    \put(20,60){\line(0,1){20}}

    \put(40,-5){\makebox(0,0)[t]{$0$}}
    \put(37,20){\makebox(0,0)[rb]{$x$}}
    \put(65,20){\makebox(0,0)[l]{$y$}}
    \put(-5,40){\makebox(0,0)[r]{$z'$}}
    \put(65,40){\makebox(0,0)[l]{$z$}}
    \put(-5,60){\makebox(0,0)[r]{$y'$}}
    \put(23,60){\makebox(0,0)[lt]{$x'$}}
    \put(20,85){\makebox(0,0)[b]{$1$}}

    \put(40,0){\circle*{3}}
    \put(40,20){\circle*{3}}
    \put(60,20){\circle*{3}}
    \put(0,40){\circle*{3}}
    \put(60,40){\circle*{3}}
    \put(0,60){\circle*{3}}
    \put(20,60){\circle*{3}}
    \put(20,80){\circle*{3}}

    \put(240,0){\line(-1,1){40}}
    \put(240,0){\line(0,1){40}}
    \put(240,0){\line(1,1){40}}
    \put(200,40){\line(0,1){20}}
    \put(200,40){\line(1,1){40}}
    \put(240,40){\line(-1,1){40}}

    \put(240,40){\line(1,1){40}}
    \put(280,40){\line(-1,1){40}}
    \put(280,40){\line(0,1){20}}
    \put(200,60){\line(0,1){20}}
    \put(280,60){\line(0,1){20}}
    \put(200,80){\line(1,1){40}}
    \put(240,80){\line(0,1){40}}
    \put(280,80){\line(-1,1){40}}

    \put(240,-5){\makebox(0,0)[t]{$0$}}
    \put(195,40){\makebox(0,0)[r]{$x$}}
    \put(240,45){\makebox(0,0)[b]{$w$}}
    \put(285,40){\makebox(0,0)[l]{$z'$}}
    \put(195,60){\makebox(0,0)[r]{$y$}}
    \put(285,60){\makebox(0,0)[l]{$y'$}}
    \put(195,80){\makebox(0,0)[r]{$z$}}
    \put(240,73){\makebox(0,0)[t]{$w'$}}
    \put(285,80){\makebox(0,0)[l]{$x'$}}
    \put(240,125){\makebox(0,0)[b]{$1$}}

    \put(240,0){\circle*{3}}
    \put(200,40){\circle*{3}}
    \put(240,40){\circle*{3}}
    \put(280,40){\circle*{3}}
    \put(200,60){\circle*{3}}
    \put(280,60){\circle*{3}}
    \put(200,80){\circle*{3}}
    \put(240,80){\circle*{3}}
    \put(280,80){\circle*{3}}
    \put(240,120){\circle*{3}}

  \end{picture}
  \caption{\hbox to3mm{\hfill}(a) Ortholattice from
\protect\cite{beran},
Fig.~9g; \hbox to15mm{\hfill} (b)
Ortholattice from \protect\cite{mphpa98}, Fig.~3.
\label{fig-transitivity}}
\end{figure}

Let us now see how a standard completeness of modeling $\cal QL$
can be proved to see where the orthomodularity in such a proof
emerges from. First we have to check whether $\equiv_i$ defines a
relation of equivalence. That the relation is symmetric, for
$\equiv_5$ it follows from $\cal QL$R4. The other four identities are
themselves asymmetric but the symmetry nevertheless holds for the
equivalence relation since that symmetry is metaimplicational---not
equational. For example, for $\equiv_1$ we prove it as
follows. From $\vdash A\equiv_1B$ by $\cal QL$R3 we get $\vdash (\neg
A\vee B)\wedge (A \vee (\neg A\wedge \neg B))$ wherefrom by
a $\cal QL$ equivalent of L8.2 we get the required result.
Note that the transitivity of $\cal QL$R2 when mapped to a lattice:
$a\equiv_ib=1\ \&\ b\equiv_ic=1\ \Rightarrow\ a\equiv_i c=1$
fail in ortholattices shown in Fig.~\ref{fig-transitivity}.
It therefore does not hold in all ortholattices as does the
relational transitivity $a=b\ \&\ b=c\ \Rightarrow\ a=c$,
but requires L8.

\bigskip

Now we can prove the following lemmas and introduce a definition.

\begin{lemma}\label{L:congruence} Relation $\approx$ defined as
\begin{eqnarray}
A\approx B\ \ {\buildrel\rm def\over =}\ \ \Gamma\vdash A\equiv_i B
\label{eq:equiv}
\end{eqnarray}
is a relation of congruence in the algebra $\cal F$.
\end{lemma}

\begin{proof}
As we have shown above, $\approx$ is an equivalence relation.
In order to be a relation of congruence, the relation of equivalence
must be compatible with the operations $\neg$ and  $\vee$.
$\vdash A\equiv_iB\ \Rightarrow\ \vdash\neg A\equiv_i\neg B$ is
nothing but $\cal QL$R3 and $\vdash A\equiv_i B\ \Rightarrow\
\vdash(A\vee C)\equiv_i(B\vee C)$ is $\cal QL$R1.
\end{proof}

\begin{definition}\label{D:equiv-class}
The equivalence class under the relation of equivalence is
defined as $|A|=\{B\in {\cal F}^\circ:A\approx B\}$ and we denote
${\cal F}^\circ/\!\approx\ =\{|A|\in {\cal F}^\circ\}$
The equivalence classes define the natural morphism
$f:{\cal F}^\circ\longrightarrow {\cal F}^\circ/\!\approx$ which gives
$f(A)\ {\buildrel\rm def\over =}\ |A|$. We write $a=f(A)$,
$b=f(A)$, etc.
\end{definition}

\begin{corrolary}\label{C:equality}
 The relation $a=b$ on ${\cal F}^\circ/\!\approx$ is given as:
\begin{eqnarray}
|A|=|B|\qquad\Leftrightarrow\qquad \Gamma\vdash A\equiv_iB.
\label{eq:equation}
\end{eqnarray}
\end{corrolary}

\begin{lemma}\label{L:lind-alg} The Lindenbaum algebra ${\cal A}=
\langle {\cal F}^\circ/\!\approx,\neg/\!\approx,
\vee/\!\approx\rangle,\wedge/\!\approx\rangle$ is a
{\em WOML}, i.e., {\em L1--L6} and {\em L8} hold for
$\neg/\!\approx$, $\vee/\!\approx$, and $\wedge/\!\approx$ 
as  $'$, $\cup$, and $\cap$, respectively.
\end{lemma}

\begin{proof}That L1--L6 hold in ${\cal A}$ is obvious.
$\cal QL$R1 gives L8.1 which together with L1--L6 gives L8
according to Theorem~\ref{th:l8.1}.
\end{proof}

We see that, as we already stressed above, to prove the
orthomodularity from $\cal QL$ in the Lindenbaum algebra the
latter need not be an orthomodular lattice. WOML suffices.
Previously we proved that the ``orthomodularity'' in $\cal QL$ is
given as: $\vdash A\vee (\neg A\wedge(A\vee B))\equiv_5 A\vee B$
is no more than an ``ortho-property'' (i.e., its lattice mapping
holds in any ortholattice). Another way of expressing orthomodularity
is $\vdash A\vee (B\wedge(\neg A\vee\neg B))\equiv_5 A\vee B$ whose
lattice mapping is nothing but L8.3, i.e., it is a WOM property
itself. This means that the  ``orthomodularity'' from $\cal QL$
sometimes maps to an ortho-property and sometimes to a
weakly orthomodular property but never to a proper orthomodular
lattice property. The following theorem explains the peculiarity.

\begin{theorem}\label{th:om}The orthomodularity lattice
property L7 holds in $\cal A$ as a consequence of the
way we define the equivalence relation in Lemma \ref{L:congruence}.
Hence, $\cal A$ is also an {\em OML}.
\end{theorem}

\begin{proof}The theorem is a direct consequence of
Theorem \ref{th:other-eq} and  rule $\cal QL$R5.
\end{proof}

As we see the orthomodularity follows from Lemma
\ref{L:congruence} for five different operations of identity
$\equiv_i$, $i=1,\dots,5$ for which in an ortholattice
Theorem \ref{th:other-eq} holds. The previous theorem is a
consequence of the very definition of the relation = between
the equivalence classes given by the Definition
\ref{D:equiv-class} as the following lemma shows.

\begin{lemma}\label{L:sixth}
There is no operation of identity $\equiv_6$ for which
$\cal QL${\em 1-6} and $\cal QL${\em R1-5} would hold, whose
lattice mapping would satisfy Eq.~{\em(\ref{eq:qm-as-id})}
in {\em WOML} and for which the orthomodularity property would
not be satisfied in $\cal A$.
\end{lemma}

\begin{proof}
Let $\cal QL${\em 1-6} and $\cal QL${\em R1-5} hold for
$\equiv_6$. We form the Lindenbaum algebra $\cal A$ for
this logic using $A\approx B\ {\buildrel\rm def\over =}\
\Gamma\vdash A\equiv_6 B$ and Lemma \ref{L:lind-alg} formulated
for this  $A\approx B$. Let us further assume that the so
obtained $\cal A$ is not orthomodular. But by
Lemma \ref{C:equality} from  $\vdash A\vee (\neg
A\wedge(A\vee B))\equiv_6 A\vee B$ which must hold in such $\cal QL$
we obtain the orthomodularity, i.e., the contradiction.
\end{proof}

The remaining lemmas and theorems we adopt from \cite{dishk}.
The first two lemmas are obvious and we omit the proofs.

\begin{lemma}\label{L:equiv-morph}\qquad\qquad $\Gamma\vdash
A\equiv_i B\qquad\Leftrightarrow \qquad f(A)=f(B)$
\end{lemma}

\begin{lemma}\label{L:model}${\cal M}\ {\buildrel\rm def\over =}\
\langle {\cal F}^\circ/\!\approx,f\rangle$ is a model of $\Gamma$.
\end{lemma}

\begin{lemma}\label{L:unary-c}\qquad\qquad $f(A)=1
\qquad\Rightarrow \qquad \Gamma\vdash A$
\end{lemma}

\begin{proof}Since $f(\neg B\vee B)=1$, the premise $f(A)=1$
yields $A\equiv_i\neg B\vee B$ by Lemma \ref{L:equiv-morph},
wherefrom by $\cal QL$R5 we get $\Gamma\vdash A$.
\end{proof}

Thus for the five above defined operations of identities---and
through them above defined equality between equivalence
classes---we obtain:

\begin{theorem}\label{th:completeness}(Completeness)
If a formula {\rm A} is true in all OML models of a set of wff's $\Gamma$,
i.e., if $f(A)=1$, then $\Gamma\vdash A$.
\end{theorem}

\begin{proof}Since $\cal M$ is a model of $\Gamma$, to be true
for $A$ in $\cal M$ means: $f(A)=1$. Hence, by the previous lemma,
we get: $\Gamma\vdash A$.
\end{proof}

The soundness of quantum logic given by Theorem
\ref{th:soundness} must be valid for OML as soon as it is valid for
WOML. Thus we obtain:

\begin{theorem}\label{th:compl-sound}
$\Gamma\vdash A$ iff $A$ is true in all {\em OML} models.
\end{theorem}

These theorems show that in the syntactical structure of  quantum
logic there is nothing orthomodular. The orthomodularity appears
through the definition of the equivalence relation. By defining
it in the standard way as above, we, in effect, introduce an
additional axiom in the lattice structure of the equivalence classes
as the Theorem \ref{th:other-eq} shows. In Sec.~\ref{sec:cl-models}
we show that one obtains an analogous result for classical logic.
The only difference will be a possibility of \{0,1\}\ evaluation
of every proposition which is not possible in quantum logic.

\subsection{Non-Orthomodular Completeness Proof for Quantum Logic}
\label{subsec:woml-complete}

As we have seen in the previous subsection the orthomodularity in
the standard completeness proof of quantum logic emerges from
nothing else but the very definition of the relation of equivalence
defined on $\cal QL$. Therefore, if we were to stay with WOML
(which served us to prove soundness) in a completeness proof for
$\cal QL$ as well, we should change the definition of $A\>\approx B$.
What we do not want in the same equivalence classes are those $A$ 
and $B$ whose lattice equality $f(A)=f(B)$ would make an ortholattice 
orthomodular when added to it. And this is exactly what O6 lattice 
offers us. Any such equality fails in it and any WOML expression 
holds in it.  

\begin{definition}\label{o6n} Letting {\em O6} represent the lattice
shown in Fig.~\ref{fig:O6}, we define ${\cal O}6$ as the set of all
mappings $o:{\cal F}^\circ\longrightarrow {\rm O}6$ such that for
$A,B\in{\cal F}^\circ$,
$o(\neg A)=o(A)'$ and $o(A\vee B)=o(A)\cup o(B)$.
\end{definition}

\begin{lemma}\label{L:congruence-noml} Relation $\approx$ defined as
\begin{eqnarray}
\!\!\!\!\!\! A\approx B\ {\buildrel\rm def\over =}\ \Gamma\vdash 
A\equiv_i B\>\&\>(\forall o\in{\cal O}6)[(\forall X\in\Gamma)(o(X)=1)
\Rightarrow o(A)=o(B)],
\label{eq:equiv-noml}
\end{eqnarray}
where $i=1,\dots,5$, is a relation of congruence in the algebra
$\cal F$.
\end{lemma}

\begin{proof}Let us first prove that $\approx$ is an equivalence
relation. $\>A\approx A\>$ and  $\>A\approx B\>\Rightarrow\>
B\approx A$ are obvious. The proof of the transitivity runs as
follows.
\begin{eqnarray}
A\approx B & \& & B\approx C\label{line1}\\
\Rightarrow\ \Gamma\vdash A\equiv_i B &\& &\Gamma\vdash
B\equiv_i C\quad\nonumber\\
&\& &\quad(\forall o\in{\cal O}6)
[(\forall X\in\Gamma)(o(X)=1)\ \Rightarrow\ o(A)=o(B)]\ \
\ \ \nonumber\\
&\& &(\forall o\in{\cal O}6)
[(\forall X\in\Gamma)(o(X)=1)\ \Rightarrow\ o(B)=o(C)]\label{line2}\\
\Rightarrow\ \Gamma\vdash A\equiv_i C &\& &\nonumber\\
(\forall o\in{\cal O}6)\!\!\!\!\!\!&[&\!\!\!\!\!\!(\forall 
X\in\Gamma)(o(X)=1)\ \Rightarrow\
o(A)=o(B)\ \&\ o(B)=o(C)].\qquad\label{line3}
\end{eqnarray}
Since all the WOML axioms and rules hold in O6, the last
metaconjunction in line \ref{line3} reduces to  $\ o(A)=o(C)\ $
by transitivity. Hence the conclusion $A\approx C$ by definition.

In order to be a relation of congruence, the relation of
equivalence must be compatible with the operations $\neg$ and
$\vee$. The proofs of the compatibilities run as follows.
\begin{eqnarray}
&& A\approx B\label{line-1}\\
&\Rightarrow &\Gamma\vdash A\equiv_i B\quad\&\quad (\forall o\in{\cal O}6)
[(\forall X\in\Gamma)(o(X)=1)\ \Rightarrow\ o(A)=o(B)]\label{line-2}\\
&\Rightarrow &\Gamma\vdash\neg A\equiv_i\neg B\ \&\ (\forall o\in{\cal O}6)
[(\forall X\in\Gamma)(o(X)=1)\ \Rightarrow\ o(A)'=o(B)']
\label{line-3}\\
&\Rightarrow &\Gamma\vdash\neg A\equiv_i\neg B\ \&\
(\forall o\in{\cal O}6)
[(\forall X\in\Gamma)(o(X)=1)\ \Rightarrow\ o(\neg A)=o(\neg B)]
\qquad\label{line-4}\\
&\Rightarrow &\neg A\approx \neg B\label{line-5}\\
\nonumber\\
&& A\approx B\label{lineV1}\\
&\Rightarrow &\Gamma\vdash A\equiv_i B\quad\&\quad (\forall o\in{\cal O}6)
[(\forall X\in\Gamma)(o(X)=1)\ \Rightarrow\ o(A)=o(B)]\label{lineV2}\\
&\Rightarrow &\Gamma\vdash(A\vee C)\equiv_i(B\vee C)\nonumber\\
&&\&\quad (\forall o\in{\cal O}6)
[(\forall X\in\Gamma)(o(X)=1)\ \Rightarrow\ o(A)\cup o(C)=o(B)\cup o(C)]
\label{lineV3}\\
&\Rightarrow &(A\vee C)\approx(B\vee C)\label{lineV4}
\end{eqnarray}
In these proofs we used $\cal QL$R3 and $\cal QL$R1 and the
corresponding lattice mappings in O6.
\end{proof}

\begin{definition}\label{D:equiv-class-sets-woml}
The equivalence class under the relation of equivalence is
defined as $|A|=\{B\in {\cal F}^\circ:A\approx B\}$ and we denote
${\cal F}^\circ/\!\approx\ =\{|A|\in {\cal F}^\circ\}$
The equivalence classes define the natural morphism
$f:{\cal F}^\circ\longrightarrow {\cal F}^\circ/\!\approx$ which gives
$f(A)\ {\buildrel\rm def\over =}\ |A|$. We write $a=f(A)$,
$b=f(A)$, etc.
\end{definition}

\begin{corrolary}\label{C:equality-non-om}
 The relation $a=b$ on ${\cal F}^\circ/\!\approx$ is given as:
\begin{eqnarray}
|A|=|B|\qquad\Leftrightarrow\qquad A\approx B
\label{eq:equation-non-om}
\end{eqnarray}
\end{corrolary}

\begin{lemma}\label{L:lind-alg-non-om} The Lindenbaum algebra
${\cal A}=\langle {\cal F}^\circ/\!\approx,\neg/\!\approx,
\vee/\!\approx\rangle,\wedge/\!\approx\rangle$ is a
{\em WOML}, i.e., {\em L1--L6} and {\em L8.2} hold for
$\neg/\!\approx$, $\vee/\!\approx$, and $\wedge/\!\approx$ 
as  $'$, $\cup$, and $\cap$, 
respectively \em [where---for simplicity---we use the same symbols
($'$ and $\cup$) as for O6 since in the paper there are no ambiguous
expressions in which the origin of the operations would not
be clear from the context].
\end{lemma}

\begin{proof}
Since all the WOML axioms and rules hold in O6 the proof follows
from the proof of Lemma \ref{L:lind-alg}.
\end{proof}

\begin{theorem}\label{th:non-om}The orthomodularity lattice
property L7 does not hold in $\cal A$.
\end{theorem}

\begin{proof}
We assume ${\cal F}^\circ$ contains at least two propositional
variables (or ``primitive'' or ``starting'' wffs). We pick
an evaluation $o$ that maps two of them, $A$ and $B$, to
distinct nodes $o(A)$ and $o(B)$ of O6 that are
neither 0 nor 1 such that $o(A)\le o(B)$
[i.e. $o(A)$ and $o(B)$ are on the same side of hexagon O6 in
Fig.~\ref{fig:O6}]. From the structure of O6 we obtain
$\>o(A)\cup o(B)=o(B)$ and $o(A)\cup(o(A)'\cap(o(A)\cup o(B)))
=o(A)\cup(o(A)'\cap o(B))= o(A)\cup 0=o(A)$. Therefore
$o(A)\cup o(B)\ne o(A)\cup (o(A)' \cap (o(A)\cup o(B))$, i.e.,
$o(A\vee B)\ne o(A\vee (\neg A\wedge(A\vee B)))$.
This falsifies $(A\vee B)\approx (A\vee(\neg A\wedge (A\vee B))$.
Therefore $a\cup b\ne a\cup (a'\cap(a\cup b))$,
providing a counterexample to the OM law for ${\cal F}^\circ/\!\approx$.
\end{proof}

Let us now reformulate the remaining lemmas and theorems from the
previous subsection.

\begin{lemma}\label{L:model-wom}$\langle{\cal F}^\circ/\!\approx,
f\rangle$ is a {\em WOML} model of $\Gamma$.
\end{lemma}

\begin{theorem}\label{th:completeness-wom}(Completeness)
If a formula {\rm A} is true in all {\rm WOML} models of a set of
wff's $\Gamma$, i.e., if $f(A)=1$, then $\Gamma\vdash A$.
\end{theorem}

\begin{proof}$f(A)=1$ is equivalent to $|A|=|B\vee\neg B|$ and
therefore to
\begin{eqnarray}
&&\Gamma\vdash A\equiv_i B\vee\neg B\quad\&\quad (\forall o\in{\cal O}6)
[(\forall X\in\Gamma)(o(X)=1)\ \Rightarrow\ o(A)=1]\qquad\\
&\Leftrightarrow&\Gamma\vdash A\quad\&\quad (\forall o\in{\cal O}6)
[(\forall X\in\Gamma)(o(X)=1)\ \Rightarrow\ o(A)=1]\\
&\Rightarrow&\Gamma\vdash A\label{last}
\end{eqnarray}
\end{proof}

\begin{theorem}\label{th:compl-sound-wom}
$\Gamma\vdash A$ iff $A$ is true in all {\em WOML} models.
\end{theorem}

\begin{proof}
Right to left metaimplication in the line \ref{last}
of Theorem \ref{th:completeness-wom} holds because
all deductions of QL are sound in WOML, and O6 is a WOML.
\end{proof}

\section{Non-Distributive Model for Classical Logic}
\label{sec:cl-models}

As in the previous section, a reader which is not at home with
methods and parlance of mathematical logic can follow this section
by reading logical expressions of the form $\vdash A$ as $a=1$ in
the lattice language.

A difference we find between classical logic and the Boolean
algebra (distributive lattice) as its standard model is that
properties that play a decisive role in the lattice do not play
such a role in the logic. And again, this is in contrast to the new
model, weakly distributive lattice which is even non-orthomodular.
To explain the difference let us consider the distributivity
property. When we add the distributivity property to an
ortholattice it becomes distributive. As in the previous section
to see what then happens in a logic we mimic logical axioms
$\vdash A$ by their lattice form $a=1$; here $a=g(A)$
and $g$ is a morphism from the logic to the lattice. Thus
$(a\wedge(b\vee c))\equiv_0((a\wedge b)\vee (a\wedge c))=1$
which we obtain by a mapping of the distributivity is true in all
weakly distributive lattices which are not even orthomodular.
However, the lattice distributivity
$(a\wedge(b\vee c))=((a\wedge b)\vee (a\wedge c))$
is true only in a distributive lattice, not in a weakly
distributive one. To understand this difference better we
briefly review a completeness proof for the standard model
in Subsection \ref{subsec:c-sound-st} and subsequently for the new
one in Subsection \ref{sec:cl-models}.

\subsection{Classical Logic}
\label{subsec:classical-logic}

Classical logic $\cal CL$  contains the connectives $\rightarrow$,
$\leftrightarrow$, $\equiv$, $\vee$, $\wedge$, and $\neg$ which we
represent with their lattice counterparts: $\rightarrow$,
$\leftrightarrow$, $\equiv$, $\cup$, $\cap$, and $'$. When we omit
parentheses, we assume these connectives bind from weakest to
strongest in this order. We also represent logical formulas, wff's,
$A$ by means of a lattice expression $a=1$ where necessary. Let
${\cal G}^\circ$ be the set of all logical expressions, i.e., well
formed formulas (wff). The latter constitutes an algebra
${\cal G}=\langle {\cal G}^\circ,\neg,\vee\rangle$.

We make use of the PM classical logical system ${\cal CL}$
[Whitehead and Russell's \it Principia Mathematica\/ \rm
axiomatization in the Hilbert and Ackermann's presentation
\cite{hilb-ack-book} (without the associativity axiom which
P.~Bernays proved redundant) but in the schemata form so
that we dispense with their rule of substitution] where
$\ A\rightarrow_0 B\ {\buildrel\rm def\over =}\>\neg A\vee B$.

\noindent
{\bf Axioms}

\medskip
{\bf $\cal CL$1.}\quad\qquad$\vdash A\vee A\rightarrow_0 A$

\smallskip
{\bf $\cal CL$2.}\quad\qquad$\vdash A\rightarrow_0 A\vee B$

\smallskip
{\bf $\cal CL$3.}\quad\qquad$\vdash A\vee B\rightarrow_0 B\vee A$

\smallskip
{\bf $\cal CL$4.}\quad\qquad$\vdash (A\rightarrow_0 B)
\rightarrow_0(C\vee A\rightarrow_0 C\vee B) $

\medskip\noindent
{\bf Rule of Inference---\it Modus ponens}

\medskip
{\bf $\cal CL$R1.} \qquad$\vdash A
\qquad\&\qquad\vdash A\rightarrow_0 B
\qquad\Rightarrow\qquad\vdash B$

\begin{definition}\label{D:delta-cl}
For $\Delta\subseteq {\cal G}^\circ$ we say $A$ is derivable from
$\Delta$ and write $\Delta\vdash A$ if there is a sequence of
formulas ended by $A$ each of which is either one of the axioms of
$\cal CL$ or is a member of $\Delta$ or is obtained from its
precursors with the help of a rule of inference of the logic.
\end{definition}

\begin{definition}\label{model-cl}We call\/\  ${\cal N}=\langle
{\cal L},h\rangle$ a model of a set of formulas  $\Delta$,
if ${\cal L}$ is a lattice {\em (WOML} or {\em OML)}
$g:{\cal G}^\circ\longrightarrow {\cal L}$ is a morphism of algebra
of wff's which satisfies $g(A)=1$ for any $A\in\Delta$;
we call the latter $A$ true in the model ${\cal N}$.
\end{definition}

\subsection{Standard Soundness and Completeness Proof for Classical Logic}
\label{subsec:c-sound-st}

The following theorem holds in $\cal CL$:
\begin{eqnarray}
\makebox[20pt]{}{\cal CL}{\bf 5.}\makebox[69pt]{}
\vdash A\vee(B\wedge C)\equiv_i (A\vee B) \wedge
(A\vee C)\makebox[69pt]{}
\label{eq:w-distr}
\end{eqnarray}
where $i=0,\dots,5$.

The theorem is usually called a distributivity law. However, when its
lattice mapping
\begin{eqnarray}
\makebox[20pt]{}{\bf L10.}\makebox[75pt]{} a\cup(b\cap c)\equiv_i
(a\cup b) \cap (a\cup c)=1\makebox[75pt]{} \label{eq:w-distr-latt}
\end{eqnarray}
is added to an ortholattice it does not make the ortholattice
even orthomodular: it does not fail in O6.
We call this property a weakly distributive one and a 
weakly orthomodular lattice to which the property is added a weakly 
distributive lattice, WDL.

We see that, as with the orthomodularity in quantum logic, in the
syntactical structure of classical logic there is nothing
distributive. The distributivity will appear as a result
of the way the relation of equivalence is usually defined in a
proof of the completeness of classical logic. To better see
this we shell first try to make $\cal CL$ complete by using the
equivalence relation given in Lemma \ref{L:congruence} instead
of the usually used one. (Note that former reduces to the latter
in a distributive algebra.) In particular, we are going to check
whether a conjecture we disproved in Theorem \ref{L:sixth} for
WOML, would perhaps work for WDL.

It is easy to verify that in $\cal CL$1-4 and $\cal CL$R1 all
expressions of the form $\vdash A$ can be written as
$\vdash A\equiv_i B\vee\neg B$. So, we can repeat the procedure
from the previous section and obtain the following theorem.

\begin{theorem}\label{th:soundness-cl}(Soundness)
If \ $\Delta\vdash A$, then $A$ is true in any {\em WDL} model.  
\end{theorem}

The critical point is definition of the equivalence relation
for the completeness proof. Standard completeness procedure
introduces it as follows.

\begin{definition}\label{D:cl-equiv-class}
The equivalence relation on ${\cal G}$ is defined as:
\begin{eqnarray}
A\approx B\qquad\Leftrightarrow\qquad \vdash A\equiv_iB,
\qquad\qquad i=1,\dots,5.
\label{eq:c-equation}
\end{eqnarray}
The equivalence class under the relation of equivalence is
defined as $|A|=\{ B\in {\cal G}:A\approx B\}$ and we denote
${\cal G}/\!\approx\ =\{|A|\in {\cal G}\}$.
\end{definition}

Only from $\cal CL$1--4 and $\cal CL$R1 we obtain:
\begin{lemma}\label{L:lind-alg-cl} The Lindenbaum algebra ${\cal B}=
\langle {\cal G}/\!\approx,\neg/\!\approx,
\vee/\!\approx\rangle,\wedge/\!\approx\rangle$ is {\em at least} a
{\em WDL}, i.e., {\em L1--L6} and {\em L10} hold for
$\neg/\!\approx$, $\vee/\!\approx$, and $\wedge/\!\approx$ 
as  $'$, $\cup$,and $\cap$, and respectively.
\end{lemma}
However, as we have shown in Subsection \ref{subsec:q-compl-st},
the very definition of the equivalence relation makes the
Lindenbaum algebra orthomodular so that we are able to prove the
following theorem.

\begin{theorem}\label{th:dist} An {\em OML} to which
{\em L10} is added is a distributive lattice.
\end{theorem}

\begin{proof} In $\cal CL$ $\ \vdash A\equiv_0B$ is equivalent to
$\ \vdash A\equiv_iB$. Therefore in WDL $\ a\equiv_0b=1$
is equivalent to $a\equiv_ib=1$. Therefore, since
Eq.~(\ref{eq:qm-id}) holds, Eq.~(\ref{eq:cl-id}) gives the
required result.
\end{proof}

Thus we end up with:
\begin{theorem}\label{th:compl-sound-cl}
$\Delta\vdash A$ iff $A$ is true in all DL models.
\end{theorem}

Had we used the following usual definition,
\begin{definition}\label{D:cl-equiv-class-usual}
The equivalence relation on ${\cal G}$ is defined as:
\begin{eqnarray}
A\approx B\qquad\Leftrightarrow\qquad \vdash A\equiv_0B,
\label{eq:c-equation-usual}
\end{eqnarray}
\end{definition}
\noindent it would make the Lindenbaum algebra $\cal B$
distributive directly by Eq.~(\ref{eq:cl-id}).

\subsection{Non-Distributive Completeness Proof for Classical Logic}
\label{subsec:wdl-complete}

Thus, we need an equivalence relation which does not
introduce orthomodularity to WDL. The following one serves the
purpose.

\begin{lemma}\label{L:congruence-nodl} Relation $\approx$
defined as
\begin{eqnarray}
A\approx B\ {\buildrel\rm def\over =}\ \Delta\vdash A\equiv_0 B
\ \&\ (\forall o\in{\cal O}6)[(\forall X\in\Delta)(o(X)=1)
\Rightarrow o(A)=o(B)],\>\quad
\label{eq:equiv-nodl}
\end{eqnarray}
is a relation of congruence in the algebra $\cal G$.
\end{lemma}

\begin{proof}
The proof actually follows from the proof of Lemma
\ref{L:congruence-noml}. We only have to prove that the rules
$\ \vdash A\equiv_0 B\ \Rightarrow\ \vdash \neg A\equiv_0 \neg B\ $
and $\ \vdash A\equiv_0 B\ \Rightarrow\ \vdash
A\vee C\equiv_0 B\vee C\ $ do hold in $\cal CL$. But this is
well known. (E.g., rules *29 and *30 on p.~116, \S 26 \cite{kleene}.)
\end{proof}

As a direct consequence of the Theorem \ref{th:non-om} we obtain
\begin{theorem}\label{th:non-d}
The Lindenbaum algebra $\cal B$ is not orthomodular and therefore
not distributive.
\end{theorem}

Hence we obtain:
\begin{theorem}\label{th:compl-sound-wcl}
$\Delta\vdash A$ iff $A$ is true in all WDL models.
\end{theorem}

\section{Conclusion}
\label{sec:conclusion}

In Sec.~\ref{sec:ql-models} we show that there are two
non-isomorphic models of the propositional calculus of quantum logic: 
an orthomodular lattice and a weakly orthomodular lattice.
In Sec.~\ref{sec:cl-models} we show that there are two
non-isomorphic models of the propositional calculus of classical 
logic: a distributive lattice (Boolean algebra) and a weakly 
distributive lattice. Hence, both calculuses are non-categorical 
and neither of them maps its syntactical structure to both its 
models. They do to one of the models and do not to the other.
Surprisingly the models which do preserve the syntactical
structure of the logics are not the standard
ones---Boolean algebra and the orthomodular lattice---but
the other ones---weakly distributive and weakly orthomodular
lattice. This immediately raises fundamental questions:
How come no one realized syntactical discrepancy between
the logics and their standard models so far? Why has the usage
of classical logic in mathematical and scientific applications
not shown contradictions? What are the repercussions for
computations and computers? \dots

As for classical logic one can answer these questions as
follows: First, very many applications have not used the
logic itself but its model instead---Boolean algebra.
Secondly, the usual two-valued logic does have only one model:
the two-element Boolean algebra---and the usual many-valued
classical logic also admits only Boolean algebra as its model.
The former claim one can easily check by means of the truth tables:
both Eq.~(\ref{eq:boole}) and Eq.~(\ref{eq:cl-id}) hold. The latter
claim can be checked in the same way, using., e.g., {\L}ukasiewicz's
three- and many-valued logic \cite{lukas} or Post's $m$-valued
logic \cite{post}. It is therefore possible that a numerical
valuation of classical logic always implies that Boolean
algebra can be the only model. In that case Eq.~(\ref{eq:boole})
would just reflect the ordering of valuation. O6 which is a 
weakly distributive model for classical logic cannot be numerically 
valuated: its left and its right nodes are not comparable, they 
are non-archimedean. Hence, the main aspect of our result is that 
the syntactical structure of classical logic corresponds to (maps to) 
the structure of the weakly distributive lattice not the one of the 
Boolean algebras. The result does not affect our usage of the 
models based on numerical valuation of classical logic but opens 
a possibility of using non-ordered lattice models which would in 
turn faithfully reflect the syntax of the logic. 

With quantum logic it is just the opposite---yes-no values cannot be 
ascribed to all quantum propositions due to the Kochen-Specker 
theorem.~\cite{peres} It is true, most applications of quantum logic 
also have not used the logic itself but its orthomodular model 
instead. Actually, what is usually called \it quantum logic\/ \rm 
in the mathematical physics literature is not the very logic but its 
orthomodular model: an orthomodular lattice itself or together with 
states defined on it.~\cite{ptak-pulm} This is because one 
straightforwardly arrives at a Hilbert space representation of quantum 
logic propositions by using its orthomodular model.~\cite{holl95} 
On the other hand, what is called quantum logic in the quantum 
computation literature is an algebra of qubits (quantum bits, 
two dimensional Hilbert space pure quantum systems) determined by 
quantum logic gates and particular algorithms (e.g., Shor's or 
Grover's).~\cite{benn-shor} However, a possible quantum logic of 
instructions for manipulating arbitrary qubits in general quantum 
computers (quantum simulators) can---in the absence of numerical 
valuation of elementary propositions---rely only on a syntactical 
structure of the qubits. ``Quantum computers require 
quantum logic, something fundamentally different to classical Boolean 
logic.'' \cite{plenio} 

Whether the ``required'' logic is the quantum logic proper (we 
considered above) or one of its two models, requires further 
investigation but certainly none of them suffices for a complete 
logic of qubits or for modeling the Hilbert space. For, a 
necessary ingredient of the latter logic is the superposition 
principle which is a property of the second order. Also one 
should define a probability measure on an orthomodular lattice 
as well as a unitary map and assume infinite-dimensionality 
if one wanted a Hilbert space description of qubits. It is a 
question whether one can simulate infinite-dimensionality by means 
of quantum logic gates of a quantum computer. Therefore in our 
newest work~\cite{mpoa99} we investigate further stronger 
than weakly orthomodular (WOM) conditions, on the one hand, and 
stronger than orthomodular (OM) ones, on the other, which Hilbertian 
lattices should satisfy. In particular, we consider generalizations 
of the so-called orthoarguesian (OA) property which when added to 
WOM lattices (WOMLs) and OM lattices OMLs) make them rich enough 
for definitions of a superposition property. A WOM OA lattice is 
then still not orthomodular (does not fail in O6) although its OA 
condition fails in all other non-OA Greechie lattices. 
(We use computer programs which do automated testing of most 
Greechie lattices with up to 14 blocks and beyond.) 

On the other hand, we will investigate whether one could use a 
finite-dimensional Hilbert space based on our OALs for qubits. 
A finite-dimensional Hilbert space allows nonstandard non-archimedean 
Keller fields in addition to the standard (real, complex, and 
quaternionic) ones. This could open a possibility for a direct 
usage of WOMOALs in the qubit logic. 

\bigskip
\bigskip
\parindent=0pt
{\large\bf Acknowledgment}

\bigskip

M.~P. acknowledges a support of the Ministry of Science of Croatia.

\parindent=20pt

\vfill\eject

\bigskip

\begin{thebibliography}{10}

\bibitem{birk-v-neum}
G.~Birkhoff and J.~{{v}on Neumann},
\newblock Ann. Math. {\bf {\bf 37}}, 823 (1936).

\bibitem{mphpa98}
M.~Pavi{\v c}i{\'c} and N.~D. Megill,
\newblock Helv. Phys. Acta {\bf {\bf 71}}, 610 (1998).

\bibitem{leblanc-book}
H.~Leblanc,
\newblock Alternatives to standard first-order semantics,
\newblock in {\em Handbook of Philosophical Logic}, edited by D.~Gabbay and
  F.~Guenthner, volume I: Elements of Classical Logic, pages 189--274,
  D.~Reidel, Dordrecht, 1983.

\bibitem{peres}
A.~Peres,
\newblock Found. Phys {\bf {\bf 26}}, 807 (1996).

\bibitem{plenio}
V.~Vedral and M.~B. Plenio,
\newblock Prog. Quant. Electron. {\bf {\bf 22}}, 1 (1998),
\newblock Retrievable at http://xxx.lanl.gov/abs/quant-ph/9802065.

\bibitem{tarski}
A.~Tarski,
\newblock Some methodological investigations on the definiability of concepts,
\newblock in {\em Logic, Semantics, Metamathematics (Papers from 1923 to 1938)
  by Alfred Tarski}, edited by J.~H. Woodger, pages 296--319, Oxford University
  Press, Oxford, 1956.

\bibitem{f-set}
A.~A. Fraenkel, Y.~Bar-Hillel, and A.~Levy,
\newblock {\em Foundations of Set Theory},
\newblock North-Holland, Amsterdam, 1973.

\bibitem{kalmb83}
G.~Kalmbach,
\newblock {\em Orthomodular Lattices},
\newblock Academic Press, London, 1983.

\bibitem{pav87}
M.~Pavi{\v c}i{\'c},
\newblock Int. J. Theor. Phys. {\bf {\bf 26}}, 845 (1987).

\bibitem{p98}
M.~Pavi{\v c}i{\'c},
\newblock Int. J. Theor. Phys. {\bf {\bf 37}}, 2099 (1998).

\bibitem{mpijtp98}
M.~Pavi{\v c}i{\'c} and N.~D. Megill,
\newblock Int. J. Theor. Phys. {\bf {\bf 37}}, 2091 (1998).

\bibitem{pav93}
M.~Pavi{\v c}i{\'c},
\newblock Int. J. Theor. Phys. {\bf {\bf 32}}, 1481 (1993).

\bibitem{beran}
L.~Beran,
\newblock {\em Orthomodular Lattices; Algebraic Approach},
\newblock D.~Reidel, Dordrecht, 1985.

\bibitem{pav89}
M.~Pavi{\v c}i{\'c},
\newblock Found. Phys. {\bf {\bf 19}}, 999 (1989).

\bibitem{dishk}
H.~Dishkant,
\newblock Rep. Math. Logic {\bf {\bf 3}}, 9 (1974).

\bibitem{dalla-c-h-b}
M.~L. {Dalla Chiara},
\newblock Quantum logic,
\newblock in {\em Handbook of Philosophical Logic}, edited by D.~Gabbay and
  F.~Guenthner, volume III., pages 427--469, D.~Reidel, Dordrecht, 1986.

\bibitem{mittelstaedt-book}
P.~Mittelstaedt,
\newblock {\em Quantum Logic},
\newblock L. M. S. Monographs; v.~18, Academic Press, London, 1983.

\bibitem{stachow-completness}
E.-W. Stachow,
\newblock J. Phil. Logic {\bf {\bf 5}}, 237 (1976).

\bibitem{harde79}
G.~M. Hardegree,
\newblock The conditional in abstract and concrete quantum logic,
\newblock in {\em The Logico-Algebraic Approach to Quantum Mechanics}, edited
  by C.~A. Hooker, volume II. Contemporary Consolidation, pages 49--108,
  D.~Reidel, Dordrecht, 1979.

\bibitem{ruettimann-book}
G.~T. R{\"u}ttimann,
\newblock {\em Logikkalk{\"u}le {d}er {Q}uantenphysik. {E}ine {A}bhandlung
  {z}ur {E}rmittlung {d}er {f}ormal-{l}ogischen {S}ysteme, {d}ie {d}er
  {n}icht-{r}elativistischen {Q}uantentheorie {z}ugrundeliegen},
\newblock Duncker \&\ Humblot, Berlin, 1977.

\bibitem{hilb-ack-book}
D.~Hilbert and W.~Ackermann,
\newblock {\em Principles of Mathematical Logic},
\newblock Chelsea, New York, 1950.

\bibitem{kleene}
S.~C. Kleene,
\newblock {\em Introduction to Metamathematics},
\newblock North-Holland, Amsterdam, 1974.

\bibitem{lukas}
J.~{\L}ukasiewicz,
\newblock {\em Selected Works},
\newblock North-Holland, Amsterdam, 1970.

\bibitem{post}
E.~L. Post,
\newblock Am. J. Math. {\bf {\bf 43}}, 163 (1921).

\bibitem{ptak-pulm}
P.~Pt\'ak and S.~Pulmannov{\'a},
\newblock {\em Quantum Structures as Quantum Logics},
\newblock Kluwer, Dordrecht, 1991.

\bibitem{holl95}
S.~S. {Holland, JR.},
\newblock Bull. Am. Math. Soc. {\bf {\bf 32}}, 205 (1995).

\bibitem{benn-shor}
A.~Barenco et~al.,
\newblock Phys. Rev. A {\bf {\bf 52}}, 3457 (1995).

\bibitem{mpoa99}
N.~D. Megill and M.~Pavi{\v c}i{\'c},
\newblock [submitted]  (1999).

\end{thebibliography}

\end{document}